\documentclass[12pt]{article}

\usepackage[english]{babel}
\usepackage[utf8x]{inputenc}
\usepackage{amsmath}
\usepackage{graphicx}
\usepackage{fullpage}
\usepackage{setspace}
\usepackage{subfigure}
\usepackage{lineno}
\usepackage{ulem}
\linenumbers

\makeatletter
\renewcommand*\env@matrix[1][\arraystretch]{%
  \edef\arraystretch{#1}%
  \hskip -\arraycolsep
  \let\@ifnextchar\new@ifnextchar
  \array{*\c@MaxMatrixCols c}}
\makeatother

\author{Jacob G. Scott \AND Anita Hjelmeland \AND Prakash Chinnaiyan \AND Alexander R. A. Anderson \AND David Basanta}
\title{PLoS CB rev1 clean}

\begin{document}

\begin{flushleft}

\Large

{\bf Microenvironmental variables must influence intrinsic phenotypic parameters of cancer stem cells to affect tumourigenicity}
\normalsize

\vspace{0.5cm}
Jacob G Scott$^{1,2*}$, Anita Hjelmeland$^{3}$, Prakash Chinnaiyan$^{4}$, Alexander R. A. Anderson$^{1}$ \& David Basanta$^{1*}$\\
\vspace{0.5cm}
$^1$ Integrated Mathematical Oncology, H. Lee Moffitt Cancer Center and Research Institute, Tampa, FL, USA\\
$^2$ Centre for Mathematical Biology, Mathematical Institute, University of Oxford, Oxford, UK\\
$^3$ Department of Cell, Developmental and Integrative Biology, University of Alabama at Birmingham, Birmingham, AL, USA\\
$^4$ Department of Radiation Oncology, H. Lee Moffitt Cancer Center and Research Institute, Tampa, FL, USA\\

\begin{abstract}
Since the discovery of tumour initiating cells (TICs) in solid tumours, studies focussing on their role in cancer initiation and progression have abounded.  The biological interrogation of these cells continues to yield volumes of information on their pro-tumourigenic behaviour, but actionable generalised conclusions have been scarce. Further, new information suggesting a dependence of tumour composition and growth on the microenvironment has yet to be studied theoretically. To address this point, we created a hybrid, discrete/continuous computational cellular automaton model of a generalised stem-cell driven tissue with a simple microenvironment.  Using the model we explored the phenotypic traits inherent to the tumour initiating cells and the effect of the microenvironment on tissue growth. We identify the regions in phenotype parameter space where TICs are able to cause a disruption in homeostasis, leading to tissue overgrowth and tumour maintenance. As our parameters and model are non-specific, they could apply to any tissue TIC and do not assume specific genetic mutations.  Targeting these phenotypic traits could represent a generalizable therapeutic strategy across cancer types.  Further, we find that the microenvironmental variable does not strongly effect the outcomes, suggesting a need for direct feedback from the microenvironment onto stem-cell behaviour in future modelling endeavours.

\end{abstract}

\vfill

$^*$Corresponding authors: jacob.g.scott@gmail.com, david@CancerEvo.org

\end{flushleft}

\newpage

\section*{Author Summary:}

\bigskip

In this paper, we present a mathematical/computational model of a tumour growing according to the canonical cancer stem-cell hypothesis with a simplified microenvironment.  We explore the parameters of this model and find good agreement between our model and other theoretical models in terms of the intrinsic cellular parameters, which are difficult to study biologically.  We find, however, disagreement between novel biological data and our model in terms of the microenvironmental changes.  We conclude that future theoretical models of stem-cell driven tumours must include specific feedback from the microenvironment onto the individual cellular behavior.  Further, we identify several cell intrinsic parameters which govern loss of homeostasis into a state of uncontrolled growth.

\newpage
\doublespace 

\section*{Introduction}
\label{sec:intro}
Heterogeneity among cancer cells within the same patient contributes to tumour growth and evolution. A subpopulation of tumour cells, called Tumour Initiating cells (TICs), or cancer stem cells, has recently been shown to be highly tumourigenic in xenograft models and have some properties of normal stem cells.  Although controversial, there is a growing body of evidence that TICs can drive tumour growth and recurrence in many cancers, including, but not limited to, brain \cite{Singh:2004fk}, breast \cite{Al-Hajj:2003uq} and colon \cite{Schepers:2012kx}.  These tumour types can be broadly classed as hierarchical tumours as they have been posited to follow some of the same hierarchical organisation as healthy stem-cell (SC) driven tissues. In these hierarchical tumors, TICs can differentiate to produce non-TIC cancer cells or self-renew to promote tumor maintenance.  As TICs have been demonstrated to be resistant to a wide variety of therapies including radiation and chemotherapy, the TIC hypothesis has important implications for patient treatments \cite{Diehn:2009vn}.  Specifically, the effect of current strategies on the tumor cell hierarchy should be defined, and TIC specific therapies are likelyto provide strong benefit for cancer patients.  

\begin{figure}[ht]
\centering
\includegraphics[scale=0.75]{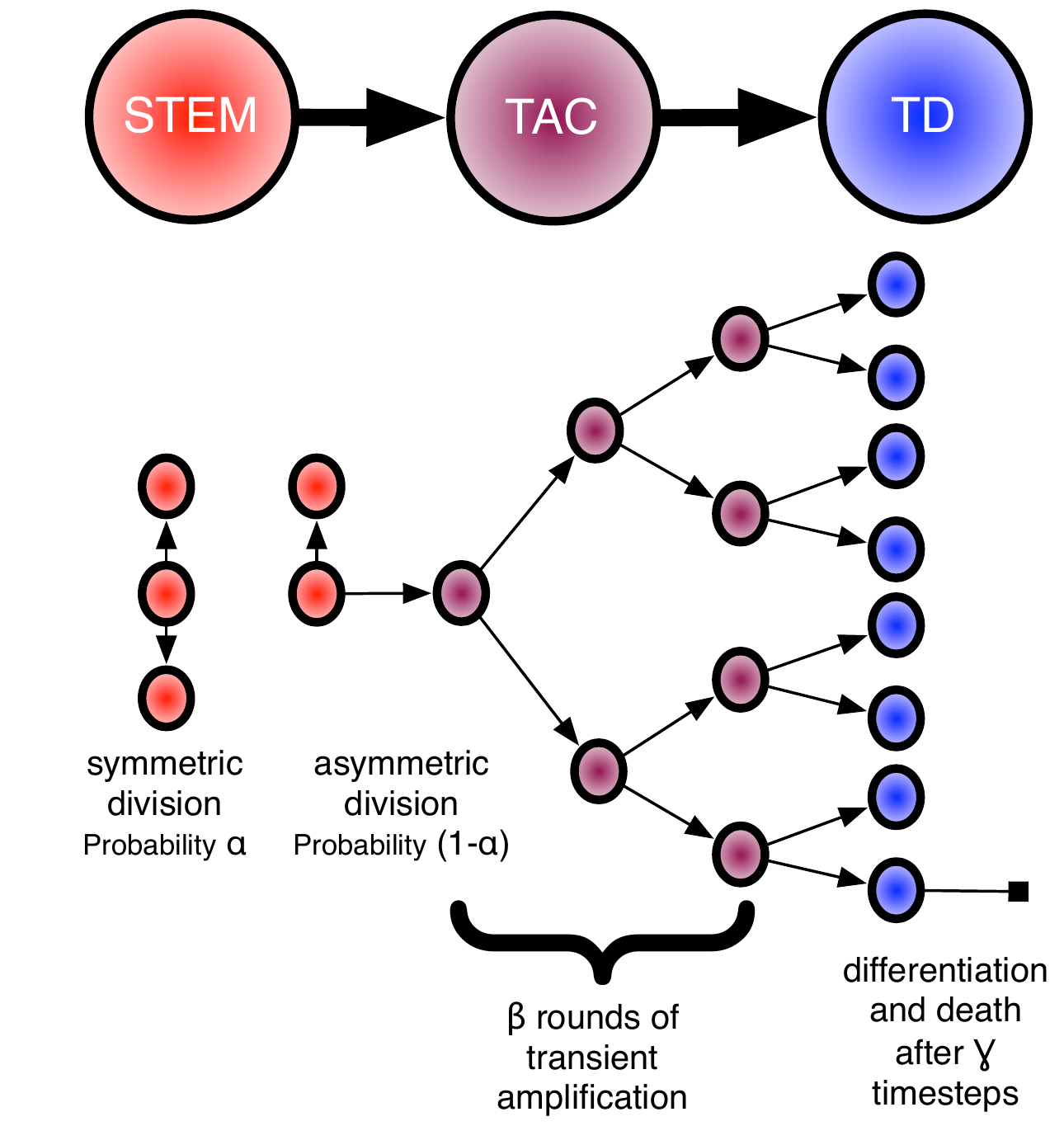}
\caption{\label{fig:HM}Cartoon representing the hierarchical model of stem-cell driven tissues. In this formulation, each stem can undergo two types of division, either symmetric (with probability $\alpha$) or asymmetric (with probability $1-\alpha$).  Each subsequently generated transient amplifying cell (TAC) can then undergo a certain number ($\beta$) of round of amplification before differentiating into a terminally differentiated cell (TD) which will live for a certain amount of time before dying ($\gamma$ timesteps).  It is these three parameters, which we assume are intrinsic to a given stem cell, which we explore in this paper.}
\end{figure}

In a simplified view of the tumour cell hierarchy, TICs can divide symmetrically or asymmetrically to, respectively, produce two TIC daughters or a TIC daughter and a more differentiated progeny \cite{Morrison:2006ys,Reya:2001zr}. More differentiated TIC progeny which still have the capability of cell division and are similar to transient amplifying cells (TACs) in the standard stem-cell model and are capable of several rounds of their own symmetric division before the amplified population then differentiates into terminally differentiated cells (TDs) which are incapable of further division.   This mode of division and differentiation, which we will call the Hierarchical Model (HM) is schematized in Fig \ref{fig:HM}.

In the HM, there are a number of cellular behaviours that govern the system. In this study, we choose to study three: the rate of symmetric versus asymmetric division of the stem cells ($\alpha$), the number of ‘rounds’ of amplification that transient amplifying cell can undergo before terminal differentiation ($\beta$), and the relative lifespan of a terminally differentiated cell ($\gamma$).  While it is a simplification of reality to study only these three parameters and leave out others (for example: differing proliferation rates for the different cell types \cite{Werner:2011fk} or the differing metabolic demands of stem vs. non-stem cells \cite{Vlashi:2011uq}) rigorous quantification of these parameters has been extremely difficult to pin down experimentally and so the majority of the work to describe them has been \textit{in silico}.  Most germane to the loss of homeostasis is the work by Enderling et al. \cite{Enderling:2009ly} which showed the changes to the size of a mutated tissue (tumour) as they varied the number of rounds of amplification of TACs.  Other recent work attempting to quantify the ratio of symmetric to asymmetric division in putative glioma stem cells was presented by Lathia et al. \cite{Lathia:2011ve}, who showed that this ratio can change depending on the medium, suggesting yet another method by which a tissue can lose or maintain homeostasis: in reaction to microenvironmental change.  A critical limitation of \textit{in vivo} lineage tracing performed to date is an inability to determine the impact of microenvironmental heterogeneity on TIC symmetric division.

While the HM appears to be quite straight forward, there is growing evidence of complexity to be further incorporated into the model.  There are likely to be differences in the extent of TIC maintenance or the ability of tumour cells to move toward a TIC state.  TICs appear to reside in distinct niches suggesting there may be differences in the biology of these cells, but defining differences in TICs is limited by cell isolation and tumour initiation methods.  Prospective isolation of TICs relies on surface markers, including CD133, CD151 and CD24 which can be transient in nature \cite{Gupta:2011qf}, due to modulation by the tumour microenvironment or methods of isolation \cite{Brescia:2012bh}. Characterisation of these sorted cells then requires functional assays including \textit{in vitro} and \textit{in vivo} limiting dilution assay as well as determination of expression of stem cell factors including Oct4, Nanog and others \cite{Hjelmeland:2011kl}.  

As the importance of TICs becomes more and more evident as it pertains to aspects of tumour progression like heterogeneity \cite{Sottoriva:2010dq}, treatment resistance \cite{Bao:2006nx,Chen:2012oq}, recurrence \cite{Dingli:2006kl} and metastasis \cite{Pang:2010tg}, the need for generalizable therapeutic strategies based on conserved motifs in these cells grows.  We therefore aim to understand how the phenotypic traits discussed earlier (asymmetric division rate, allowed rounds of transient amplification and lifespan of terminally differentiated cells) and microenvironmental changes (modelled as differences in oxygen supply) effect resultant tissue growth characteristics.

To this end, we present a minimal spatial, hybrid-discrete/continuous mathematical model of a hierarchical SC-driven tissue architecture which we have used to explore the intrinsic, phenotypic, factors involved in the growth of TIC-driven tumours.  We consider parameters that involve the rates of division of the cells involved in the hierarchical cascade as well as micro-environmental factors including space and competition between cell types for oxygen.  We present results suggesting that there are discrete regimes in the intrinsic cellular parameter space which allow for disparate growth characteristics of the resulting tumours, specifically: TICs that are incapable of forming tumours, TICs that are capable of forming only small colonies (spheres), and TICs that are capable of forming fully invasive tumours \textit{in silico}, just as we see diversity in biological experiments (Fig \ref{fig:exemplar}).

\section*{Methods}
\label{sec:methods}

\begin{figure}[ht]
\centering
\includegraphics[scale=0.25]{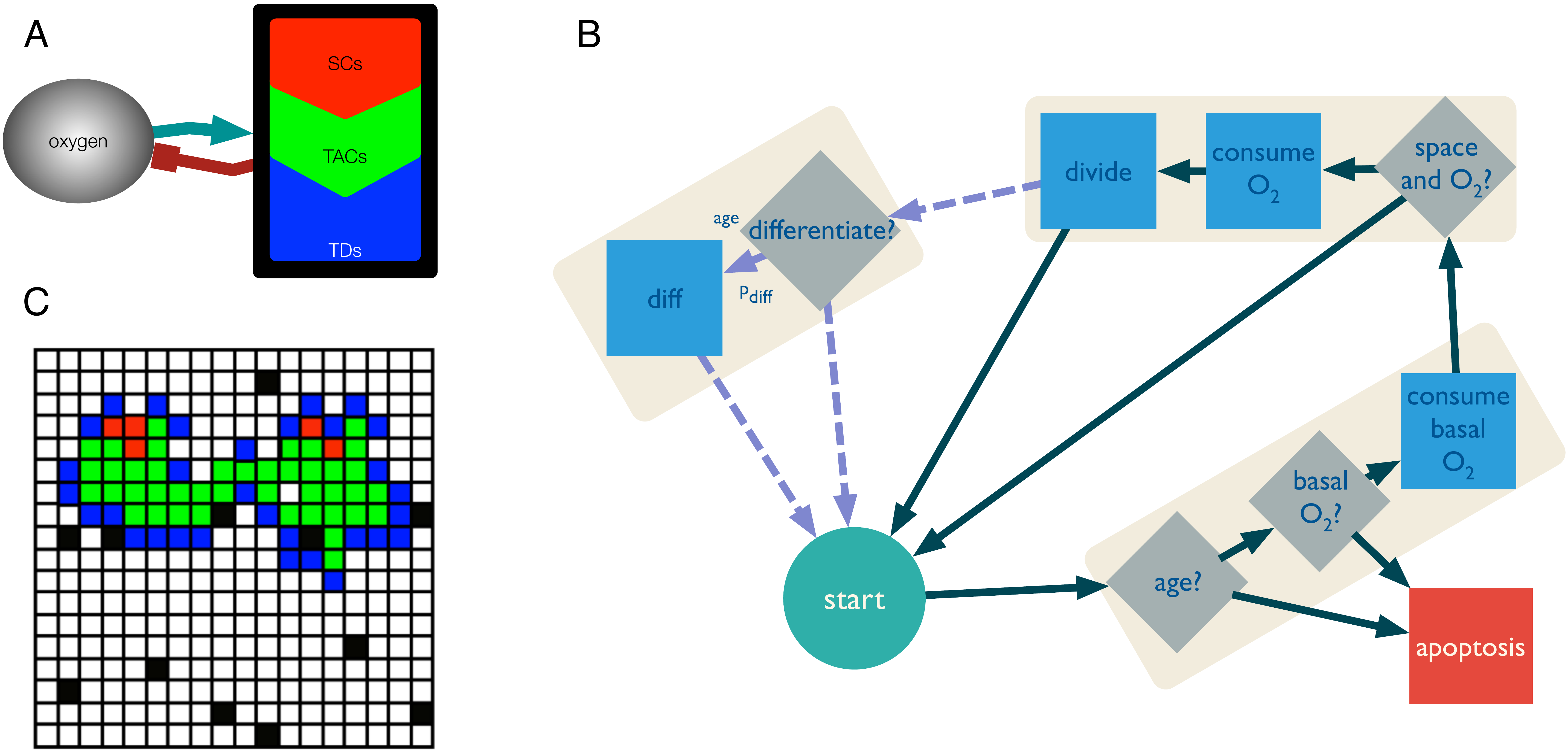}
\caption{Computational model description. (A) The model includes three different cell types: stem, progenitor and differentiated. All cell types interact with the microenvironment in the form of oxygen tension. (B) The behaviour of each cell type is captured by a flowchart. The last segment with  discontinuous arrows represents behaviour that is specific to the stem cells. (C) The cells are represented as agents inhabiting points in a grid in a 2D space with 500x500 grid points. Stem cells are represented as red points, progenitor as green and fully differentiated as blue. The vasculature is represented as oxygen source points in black.}
\label{fig:CAcartoon}
\end{figure}

Our model is based on a hybrid, discrete-continuous cellular automaton model (HCA) of a hierarchically structured tissue.  HCA models have been used to study cancer progression and evolutionary dynamics since they can integrate biological parameters and produce predictions affecting different spatial and time scales \cite{Sottoriva:2010dq,Anderson:2005hc,Anderson:2009bs,Anderson:2006ij,Basanta:2009fv, Sottoriva:2010cr}. As shown in figure \ref{fig:CAcartoon}C, cells are modelled in a discrete fashion on a 500x500 2-D lattice. This comprises approximately $1cm^2$ where we assume a cell diameter of 20 micrometers \cite{Melicow:1982dz}.  The domain has periodic boundary conditions but the simulations are stopped when a cell reaches one of the boundaries. Every time step, cells are iterated in a random fashion as to avoid any bias in the way that cells are chosen. Figure \ref{fig:CAcartoon}A shows that, although all cells are assumed to have the same size and shape, they can only be one of three different phenotypes: TICs capable of infinite divisions, TACs which are capable of division into two daughters for a certain number ($\beta$) of generations, and TDs which cannot divide but live and consume nutrients for a specified lifetime ($\gamma$).  Modes of division for TICs include asymmetric division (with probability $1-\alpha$), which is division into one TIC daughter  and one TAC daughter and symmetric division, which is division into two TIC daughters (probability $\alpha$).

The continuous portion of this model is made of up the distribution and consumption of nutrients (in this case modelled only as oxygen).  Vessels, which are modelled as point sources and take up one lattice point (V$_{i,j}$ in Equation \ref{eq:1}), are placed randomly throughout the grid at the intiation of a given simulation, in a specified density ($\Theta$).  Each of these vessels supplies oxygen at a constant rate ($\lambda$) which then diffuses into the surrounding tissue.  The diffusion speed/distance is described by Equation \ref{eq:1}, where $O(x,y,t)$ is the concentration of oxygen at a given time ($t$), and place ($x,y$), $D_O$ is the diffusion coefficient of oxygen, $\lambda$ is the rate of oxygen production from a blood vessel, $\mu_s$, $\mu_p$, and $\mu_t$ are the rates at which TIC, TAC and TD cells consume oxygen. The difference in time scales that govern the diffusion of nutrients and that at which cells operate is managed by updating the continuous part of the model 100 times per time step. During each update the oxygen tension in a given grid point is updated with the values of the surrounding cells using a von Neumann neighbourhood modulated by the diffusionrate ($D_O$).

\begin{equation}
\frac{\partial O(x,y,t)}{\partial t} = D_O \nabla ^2 O(x,y,t) + \lambda
V_{x,y} - \mu_S S_{x,y} - \mu_P P_{x,y} - \mu_T T_{x,y} \label{eq:1}
\end{equation}

\begin{figure}[ht]
\centering
\includegraphics[scale=1]{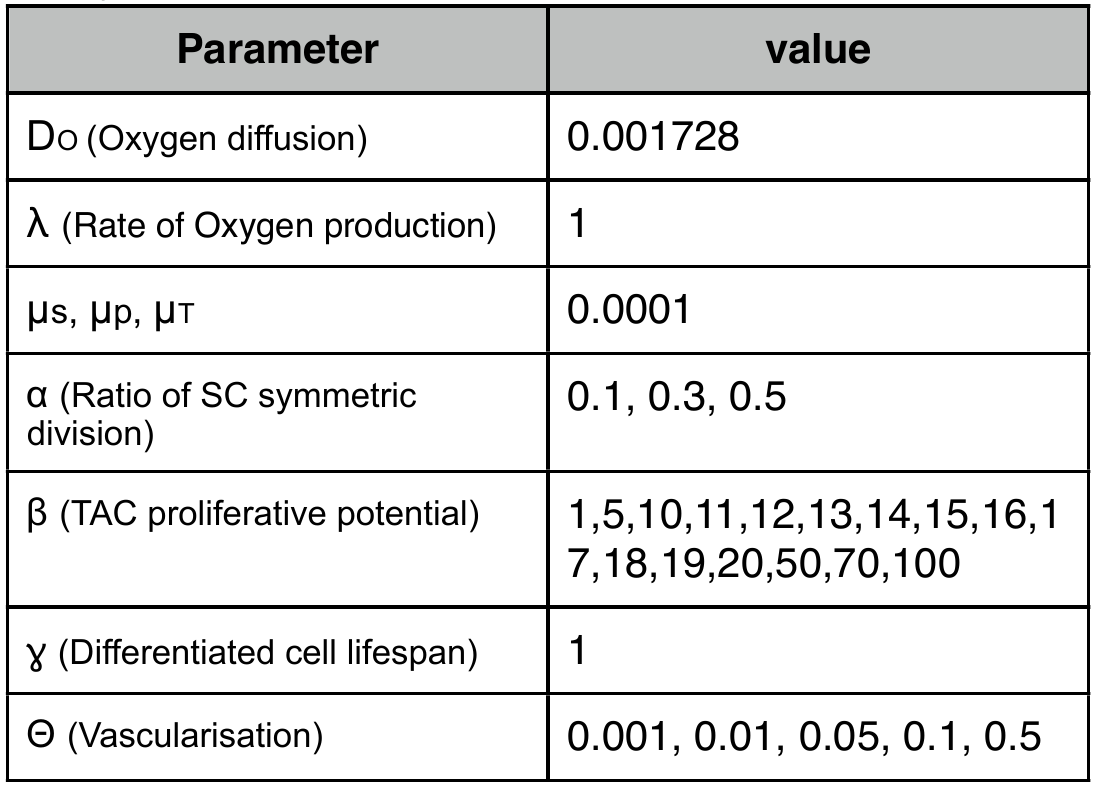}
\caption{Model parameters.}
\label{fig:table1}
\end{figure}
Any simulation performed by this model can be characterised by the parameters found in table 1. The most relevant parameters for the question we are trying to address are the following: 
\begin{itemize}
\item Symmetric/asymmetric division rate of stem cells ($\alpha$)
\item Vascular density of the tissue ($\Theta$)
\item Number of allowed divisions of TACs ($\beta$)
\item Lifespan of TDs ($\gamma$)
\end{itemize}
In each case, as can be seen in figure \ref{fig:CAcartoon}, a simulation is seeded with one TIC with a given set of intrinsic parameters ($\alpha$, $\beta$, $\gamma$) governing its and its progeny’s behaviour, which is placed in the centre of the computational domain. The domain is initialised with as many randomly placed oxygen source points (vasculature) as described by the vascular density parameter ($\Theta$).

\section*{Results}

A systematic parameter exploration of the three key parameters relating to vascularisation of the domain, symmetric vs. asymmetric division ($\alpha$) and progenitor division potential ($\beta$) was performed. We also explored the parameter determining the lifespan of differentiated cells ($\gamma$) and found that the only impact of longer lifespans is an increase in the amount of time before the simulations reach a steady state, but does not change the qualitative nature of the results. These results are summarised in Figure 4. Each of the three panels represents the results for a different degree of vascularisation (0.01, 0.05 and 0.1). A density of vascularisation of 0.05 would mean 12,500 oxygen sources in the domain. To determine the diffusion coefficient, we used the estimate of approximately 70 micrometers of effective oxygenation \cite{Hall:2012fu}.  Each plot shows the total tissue size after 50,000 time steps as we change the proliferative potential of progenitor cells. Each of the lines shows a different ratio of symmetric vs asymmetric divisions. These results show that all these three parameters have a critical range where homeostasis is disrupted (tumourigenesis). 
\begin{figure}[ht]
\centering
\includegraphics[scale=0.50]{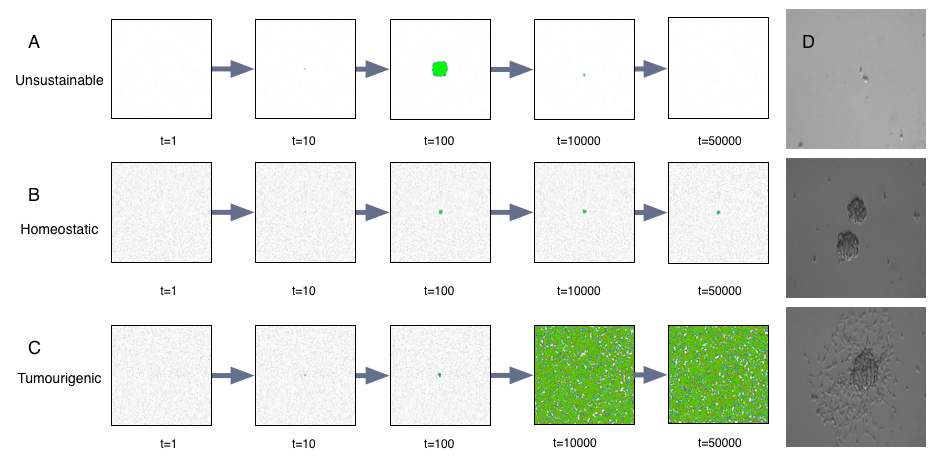}
\caption{Three different examples of simulations resulting from the 
computational model. Each simulation represents one of the typical outcomes.  Each begins with a single TIC seeded in the middle of the computational domain and the same vascular density. In each situation the phenotype parameters are slightly different, resulting in (A) An unsustainable tissue (parameters: $\Theta=0.001$, $\alpha= 0.3$, $\beta=50$ and $\gamma=1$ day), (B) A homeostatic tissue where the balance of stem cell sell renewal and progenitor proliferation leads to a tissue whose overall size remains relatively constant over time, possibly representing a dormant tumor (parameters: $\Theta=0.05$, $\alpha= 0.3$, $\beta=15$ and $\gamma=1$ day) and, (C) Neoplastic-like tissue where the tissue overgrows the computational domain (parameters: $\Theta=0.05$, $\alpha= 0.3$, $\beta=5$ and $\gamma=1$ day). (D) Bright field images of clonal CD133+ patient derived glioblastoma cell lines cultured in Neurobasal supplemented with EGF, FGF and B27, exhibiting similar phenotypic variability to the computation model outcomes.}
\label{fig:exemplar}
\end{figure}

Figure \ref{fig:exemplar} shows examples of the typical results produced by this model. Although the proliferation rates of all the cells remain the same, due to space constraints and the differences in $\alpha$, the population of TICs does not grow at the same rate as the non-stem population. Figure \ref{fig:exemplar}A shows an example of an unviable tissue (parameters: $\Theta=0.001$, $\alpha= 0.3$, $\beta=50$ and $\gamma=1$ day) where the vascularisation does not support the potential tissue size of that TIC, resulting in an area of hypoxia affecting the region that contains the TIC. That leads to the death of the stem cell and, eventually, the rest of the cells in the tissue. Figure \ref{fig:exemplar}B shows a case of slightly increased symmetric division, resulting in a dynamic homeostasis where cell birth and death is balanced so that tissue size remains relatively constant - which could represent the enigmatic dormant phase \cite{Enderling:2009ly}. Finally, figure \ref{fig:exemplar}C shows an example where the system never achieves true homeostasis. In this case $\alpha$ is slightly higher when compared with the previous example, suggesting a critical value at which overgrowth occurs. Over time, the number of TICs increases, allowing for the `tumour phenotype': unconstrained growth. Although this leads to areas of hypoxia, cells survive in the periphery of the blood vessels and keep growing until they take over the entire domain.

Unsurprisingly, the higher the vascularisation of the domain the greater the tissue size it can support.  Past a certain threshold, however, the difference becomes negligible and more remarkably, the qualitative dynamics are unchanged by any change in the microenvironment. The same effect is evident in the other two parameters, the ratio of symmetric vs asymmetric division ($\alpha$) of TICs and the proliferative potential of TACs ($\beta$). Regardless of the vascularisation, disruption of homeostasis only occurs when the proliferative potential of TACs ($\beta$) is below a maximum value of about 15. For values of symmetric division ($\alpha$) above 0.3, the values for $\beta$ in which this overgrowth occurs becomes even more restrictive with a range of approximately 10-15.

Interestingly, we observed a conserved decrease in overall tissue size for the highest value of symmetric division, $\alpha=0.5$, when the progenitor cells were allowed only 5 divisions ($\beta=5$).  We believe this phenomenon represents a situation where the tissue is not able to grow to its potential as the stem cells themselves occupy too much space, and never allow the progenitors to contribute as much as they could to the overall population.  This is a supposition however, and deserves closer study.

\begin{figure}[ht]
\centering
\subfigure
{\includegraphics[width=5.4cm,height=5cm]{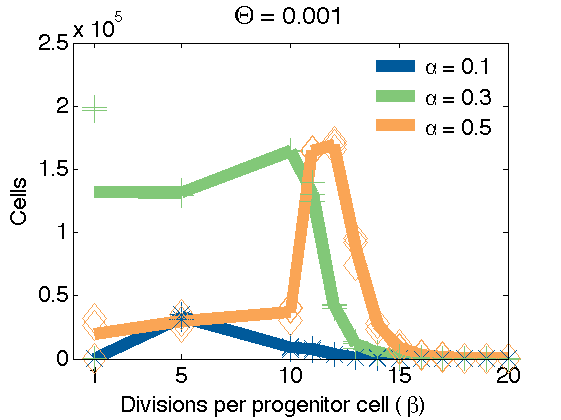}\label{fig:summar001}}
\subfigure
{\includegraphics[width=5.4cm,height=5cm]{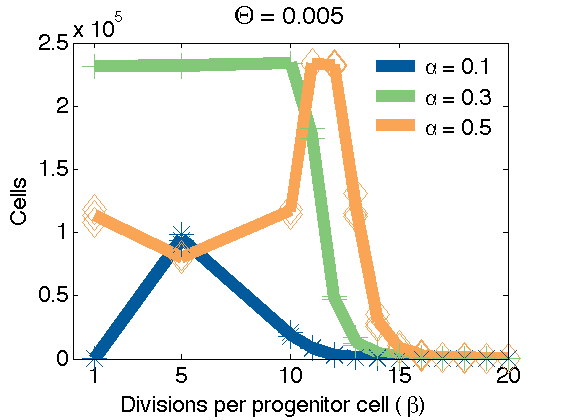}\label{fig:summar005}}
\subfigure
{\includegraphics[width=5.4cm,height=5cm]{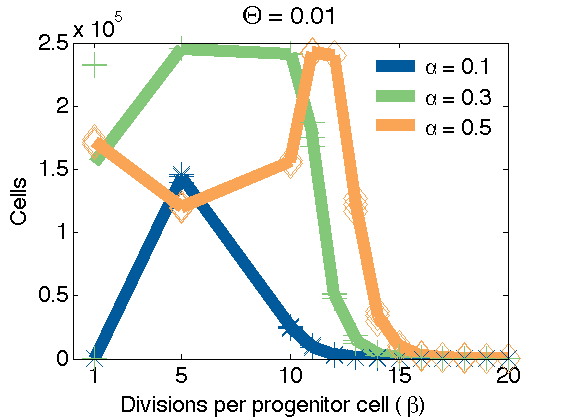}\label{fig:summar01}}
\label{fig:result}
\caption{Size of tissues vs. progenitor proliferative potential achieved by simulations using different levels of vascularisation and ratios of symmetric vs asymmetric divisions. Lines represent averages for each of the three realisations in each scenario. (Left). Low vascularisation density of 0.01 (Centre) Normal vascularisation density of 0.05 (Right) High vascularisation density of 0.1. In each of these cases, the maximum tissue size will depend on the right combination of $\alpha$ and $\beta$.}
\end{figure}
%
%

Of note as well: in no simulation did we observe the `tumour phenotype' for a value of $\alpha<0.3$, suggesting something akin to a `phenotypic tumour suppressor' function for this parameter.  As observed biologically \cite{Lathia:2011ve}, this ratio is highly susceptible to changes in microenvironment, suggesting an extension of this minimal model to include the microenvironmental factors measured in that study.  How to incorporate the changes observed in that study into a mechanistic HCA model however, is not trivial, and we reserve it for a future extension of this work.

\section*{Discussion}
In this paper we have presented a simple computational model of the HM of a TIC-driven tissue.  Our results show that there are distinct regions in parameter space (that directly correlate to the intrinsic TIC phenotype space) that encode vastly different behaviour in the tissue (or tumour) arising from the TIC in question.  These parameters represent different TIC phenotypes, and therefore do not represent any specific genetic mutation.  In this way, we hope to generalise the intrinsic alterations which a TIC could undergo much in the same way that the ‘hallmarks of cancer’ have generalised non TIC-specific alterations \cite{Hanahan:2011qa}: our end goal being the identification of treatment strategies to target these phenotypes to slow or stop the progression of TIC-driven cancers.

Because of the difficulties in understanding TIC specific traits \textit{in vivo}, the biological data to support these conclusions remains sparse.  There have been some carefully undertaken \textit{in vitro} experiments on single TICs in glioblastoma, a highly invasive and malignant brain tumour, which suggest that TIC specific division behaviour ($\alpha$ in our model) is variable and changes based on environmental cues \cite{Lathia:2011ve}.  Further work has shown that the other microenvironmental cues, such as acidity \cite{Hjelmeland:2011kl} and hypoxia \cite{Heddleston:2009mi,Li:2009pi,Seidel:2010ff,Bar:2010lh,Soeda:2009fu,Mathieu:2011ye,Kolenda:2011qo} can also alter the prevalence of the stem phenotype by utilising functional markers of stemness, but the mechanism for this increase is, as of yet, imperfectly understood.  

Of note, our simulations do not show a significant TIC population dependence on vascular density ($\Theta$), a surrogate for hypoxia, or a change in stem composition (see supplemental spreadsheet), suggesting a flaw in the model.  To rectify this, future iterations of this model should include direct feedback onto the cellular parameters from the microenvironment.  We aim to parameterize this dependence by specific \textit{in vitro} experiments designed to \textit{quantify this effect}, rather than just elucidate its existence.  Other future developments of this model should take into consideration the emerging body of work suggesting that the proportion of TICs within a tumour is directly affected by therapy and not just physiologic growth factor controls \cite{Vermeulen:2010il}.  There is now evidence in several cancers to suggest that radiation increases the size of the TIC pool.  Specifically, in breast cancer, it has been shown that radiation therapy induces non-stem cancer cells to de-differentiate into TICs \cite{Lagadec:2012tw}. Further, experimental studies have shown radiation increases the TIC pool in glioblastoma \cite{Tamura:2010gb}, which has often been attributed to radiation resistance \cite{Bao:2006nx} alone. A new study by Gao et al. \cite{Gao:2013mb}, however, has shown \textit{in silico} and \textit{in vitro} that radiation can effect the symmetric to asymmetric division ratio (our intrinsic parameter $\alpha$), yielding further clues about the mechanism of this TIC pool expansion.

This behaviour, dedifferentiation due to treatment related microenvironmental factors, has not yet been considered in any spatial theoretical models.  Dedifferentiation due to `niche' specific factors was studied by Sottoriva et al. \cite{Sottoriva:2010cr}, whose findings were similar to ours: that the microenvironment made no significant change to the overall tumour growth dynamics. Beyond this single spatial study, the concept of SC dedifferentiation is gaining more and more attention in conceptual theoretical treatments \cite{Vermeulen:2012kh} and has been modelled with a deterministic ordinary differential equation system for a well-mixed population of cells \cite{Leder:2010jl}.  

We, as well as others, find that the HM of tissue growth does not completely capture all the necessary dynamics that characterise cancer growth - but there is still a great deal of understanding to be gained from studying this formalism.  To this end, we have performed a study of the factors related to TICs driving this dynamic and have identified several key factors which promote increased growth of the resultant tumour.  In the same way that Hanahan and Weinberg \cite{Hanahan:2011qa} have simplified the myriad (epi)genetic alterations which a tumour can undergo into the ‘hallmarks of cancer’ we seek to distill the traits of TICs in the same way.  Specifically, we have found that the number of allowed divisions of TACs exhibits bounds outside of which tumour growth is unsustainable.  This finding has been corroborated independently by recent work from Morton and colleagues \cite{Morton:2011cq}.  Further, there is a specific balance of symmetric to asymmetric division which keeps tumours from overgrowing; almost acting as a phenotypic ‘tumour suppressor’.  Indeed, changes in this ratio have been recently hypothesized to underlie the increasing stem pool in glioblastoma after irradiation \cite{Gao:2013mb}, and could also represent a key to understanding tumour dormancy \cite{Enderling:2009ly}.

In summary, we have presented a minimal spatial Hybrid Cellular Automaton model of the HM of a TIC-driven tissue in which we have explored generalised TIC phenotypic traits and have identified several key cellular parameters which influence the overall tissue behaviour.  While our model does capture a number of salient phenotypic characteristics of TICs that seem to be conserved, it fails to capture the recently observed changes in stem fraction secondary to microenvironmental perturbations.  This is an indication that any computational model of a stem-hierarchical tissue, or tumour, built from this point on must not only include the physical microenvironment, but also feedback from the microenvironment onto the specific cellular parameters encoded in the HM.

Therefore, this endeavour has identified the crucial point that the microenvironment must effect the behaviour of the cells within the HM, and also several conserved phenotypic ‘hallmarks’, which could be the result of any number of (epi)genetic alterations or microenvironmental perturbations.  By focussing on phenotype instead of genotype, and identifying the key points of the HM of stem-cell driven tumour growth, we have provided a beginning to identification of the therapeutic targets to a more tractable set as compared to the panoply of possible mutations encoding similar traits.  Only with this sort of distillation of the biological complexity inherent to cancer initiation (and indeed progression) can we hope to make progress against this disease. 

\singlespace
\bibliography{main.bib}
\bibliographystyle{unsrt}

\end{document}